\documentclass[aip,graphicx]{revtex4-1}
\usepackage{graphicx}
\usepackage{amsmath, amsfonts, amssymb}
\usepackage{xcolor}
\usepackage{subcaption}
\draft 

\begin{document}


\title{Merging of zonal flows in gyrofluid resistive drift-wave turbulence} 



\author{Fabian Grander}
\email{Fabian.Grander@uibk.ac.at}
\affiliation{University of Innsbruck; Department for Ion physics and Applied physics}
\author{Tobias Gröfler}
\affiliation{University of Innsbruck; Department for Ion physics and Applied physics}
\author{Franz Ferdinand Locker}
\affiliation{University of Innsbruck; Department for Ion physics and Applied physics}
\author{Manuel Rinner}
\affiliation{University of Innsbruck; Department for Ion physics and Applied physics}
\author{Alexander Kendl}
\affiliation{University of Innsbruck; Department for Ion physics and Applied physics}

\date{\today}

\begin{abstract}
    {The following article has been submitted to Physics of Plasmas. After it is published, it will be found at \url{https://pubs.aip.org/aip/pop}. Copyright (2026) Fabian Grander. This article is distributed under a Creative Commons Attribution (CC BY) License.

    Abstract:
    Non-linear dynamics of zonal flows is investigated in the context of the gyrofluid modified Hasegawa-Wakatani model. Merging of zonal flows and the chaotic developement of the initial zonal flow pattern is explored. Conservation equations for zonal flow momentum and energy with consistent finite Larmor radius (FLR) effects are derived and used for a quantitative analysis of zonal flow mergers in numerical simulations. The nonlinear local Reynolds stress transfer as opposed to (hyper)viscous dissipation is found to be the main cause of merging. The applicability of the concept of a ”phase transition” in the strict thermodynamical sense is discussed in context of zonal flow transition hysteresis.}
\end{abstract}


\maketitle 

\section{Introduction}\label{sec:intro}
Understanding turbulence and instabilities in magnetized plasmas\cite{Scott1, Scott2} is foundational for research on magnetically confined plasmas e.g. in fusion experiments. It remains an active field of research even in highly simplified "toy-models". The delta-f, isothermal Hasegawa-Wakatani\cite{hasegawa1983edge} model is the most basic system to study self-sustained, gradient-driven, quasi two-dimensional\cite{kraichnan1980two, tabeling2002two, kendl2008two, gurcan2023structure}, resistive drift-wave turbulence\cite{hasegawa1979nonlinear, camargo1995resistive, horton1999drift, scott2002nonlinear}. It can be modified to contain the interaction of turbulence and zonal flows\cite{chen2000excitation, smolyakov2000coherent, numata2007bifurcation, pushkarev2013zonal, gurcan2015zonal, zhu2020dimits, singh2021unified,  guillon2025phase, ivanov2025suppression, sasaki2026bifurcation} and to include consistent \textsc{flr} effects within the gyrofluid\cite{hammett1992fluid, dorland1993gyrofluid, kendl2018vortex, grander2024hysteresis, kendl2025ghw} framework.  

The gyrofluid modified Hasegawa-Wakatani (\textsc{gmhw}) model can be derived from higher fidelity models. One option is by reducing 3D delta-f gyrofluid models\cite{scott2003computation} using the parallel Hasegawa-Wakatani closure. Another way to derive the \textsc{gmhw} model is to take the delta-f limit on 2D full-f models\cite{madsen2013full, held2020pade, wiesenberger2024effects}.

The transition of turbulence to zonal flows in the context of the modified Hasegawa-Wakatani model\cite{numata2007bifurcation} has been investigated over two decades. Recently it was observed in Ref. \citenum{grander2024hysteresis} that zonal flows do not necessarily remain stable but undergo changes even in apparent equilibrium states in long-time numerical simulations. These changes appear in the form of the "merging" of zonal flow streams. This is in line with the observations presented in Ref. \citenum{ivanov2020zonally} in the context of ITG drift-fluid simulations. 


In Ref. \citenum{grander2024hysteresis} it was demonstrated that the resistive drift-wave turbulence to zonal flow (forward and backward) transitions show aspects of a hysteresis loop. In Ref. \citenum{guillon2025phase} it was suggested that this hysteresis is a feature of a phase transition\cite{guillon2025phase, guillon2025flux, guillon2025self}.

In this article we further investigate the nonlinear dynamics of zonal flows in the context of the gyrofluid modified Hasegawa-Waktani model during onset and merging of zonal flows. The initial zonal flow patterns develop chaotically in the sense that they are sensitive to (small) changes in the initial conditions.
Nonlinear ("chaotic") local momentum transfer through the Reynolds stress appears to be a relevant ingredient for merging of zonal flows. This nonlinear behavior is discussed in context of the question if turbulent zonal flow transitions in the Hasegawa-Wakatani system could be identified as a phase transition in the strict thermodynamical sense.

To this end the hysteresis loop in the transition was reproduced in a numerical experiment. Also it was extended to include finite Larmor radius (\textsc{flr}) effects. Non-linear effects in the hysteresis loop speak against calling it a phase transition in the thermodynamic sense.

The rest of this article is organized as follows. First we give the model equations and some details about their numerical implementation in section \ref{sec:model_and_code}. In section \ref{sec:zonal flow-theory} we present the derivation of the zonal flow momentum  and zonal flow energy conservation equations respectively from the model given in section \ref{sec:model_and_code}. 

Section \ref{sec:why_zf_merge} is the main part of the paper. The roles of sensitivity to initial conditions, (nonlinear) local Reynolds stress, and (linear) hyperviscosity for zonal flow momentum evolution are explored. 

In the rest of the paper some implications of this nonlinear behavior of zonal flows are explored. 
In section \ref{sec:equilibrium-state} we investigate the possibility of equilibrium states, when zonal flows are present, since it cannot be predicted what the initial zonal flow state is or if flows will merge respectively.
In the final section \ref{sec:flr-and-stuff} we show results of a parameter scan over intermediate values of the parallel-coupling parameter (or: adiabaticity parameter\cite{guillon2025phase}) $C \in [0.3,2]$ and the ion to electron temperature ratio $\tau \in [0,2]$ on dominant radial modes. The parallel coupling parameter $C$ controls the phase shift of the electrostatic potential $\phi$ and the electron gyrocenter density $N_e$ due to electron resistivity along the constant, static magnetic field. The ion to electron temperature ratio $\tau$ is a key figure for the strength of \textsc{flr} effects.   In this section we also present the reproduction of the hysteresis loop mentioned above and discuss the validity of the term \textit{phase transition} from thermodynamics in this context.

\section{Model equations, code and simulation parameters}\label{sec:model_and_code}

    \subsection{The model equations}
    The  \textsc{gmhw} model equations are presented in this section. The equations are given in 2D slab-geometry, where $x$ corresponds to the radial and $y$ corresponds to the poloidal direction respectively. One of the underlying assumptions is a strong, constant and static background magnetic field $B$, that determines the toroidal or parallel direction. The normalized delta-f model equations are 
    \begin{align}
        \partial_t N_e + \frac{1}{\delta} [\phi, N_e] - g_n \partial_y \phi = C (\hat{\phi} - \hat{N}_e) - \nu \nabla^4 N_e, \label{cont_ne}\\
        \partial_t N_i + \frac{1}{\delta} [\phi_i, N_i] - g_n \partial_y \phi_i= - \nu \nabla^4 N_i,\label{cont_ni}\\
        \frac{1}{\tau} (\Gamma_0 - 1) \phi = N_e - \Gamma_1 N_i\label{pol}.
    \end{align}

    Equations \eqref{cont_ne} and \eqref{cont_ni} evolve the  fluctuations of the \emph{gyrocenter} densities for electrons $N_e$ and ions $N_i$. The full quantities in the delta-f approximation are denoted as $\Bar{N}_s = N_s + N_0(x)$ where $N_0 \gg N_s$ for both species $s \in (e,i)$.
    
    The nonlinear advection of the fluctuations is expressed with the Poisson brackets. They are weighed with the scale-ratio $\delta \equiv \rho_0/L_\perp$ where $\rho_0 \equiv \sqrt{m_i T_e}/(eB)$ denotes the drift scale ($m_i$ is the ion mass, $T_e$ the electron temperature, $e$ the unit-charge and $B$ the constant, static background magnetic field strength) and a perpendicular scale $L_\perp$, which is set as the density gradient length. They express the nonlinear $E\times B$ advection with $\phi$ and $\phi_i\equiv \Gamma_1 \phi$ acting as stream-functions respectively (the gyro-operators $\Gamma_1$ and $\Gamma_0$ are introduced below). In addition to that, the advective derivative also acts on the static background density $N_0(x)$ with a constant, radial gradient where $g_n \equiv \delta^{-1} \partial_x N_0(x)$.

    The right hand side of equation \eqref{cont_ne} contains the term with the parallel coupling parameter (or: adiabaticity parameter\cite{guillon2025phase}) $C \equiv (L_\perp/ c_0) k_\parallel^2 T_e/(\eta e^2 n_0)$ where $c_0 \equiv \sqrt{T_e/m_i}$ is the thermal speed, $k_\parallel$ a dominant parallel wave-number and $\eta$ the resistivity. 
    This term includes the "modified" form of the Hasegawa-Wakatani equations, where the difference of the non-zonal parts of the Reynolds decomposed\cite{held2018non} density $\hat{N}_e \equiv N_e - \langle N_e \rangle_y$ and potential $\hat{\phi} \equiv \phi - \langle \phi \rangle_y$ is calculated. In the 2D slab geometry of this model the average over the $y$-direction $ \langle - \rangle_y = \frac{1}{L_y} \int_0^{L_y} dy$ (with the $y$-domain-length $L_y$) expresses the zonal average. 
    
    Finally both of the continuity equations include a term containing the hyperviscosity\cite{scott1988implicit, scott1992mechanism, smith1997eddy} parameter $\nu$. This term has no physical meaning, however it is necessary to ensure numerical stability via regularization. It is there to quench numerical artifacts on the grid level, that would pollute energy-containing scales through the turbulent cascade\cite{Scott1}. 
    Ideally, it should not interfere with large-scale physics.
    
    The density-continuity equations  \eqref{cont_ne} and \eqref{cont_ni} are closed with the polarization equation \eqref{pol}, which governs the evolution of the electric potential $\phi$ and thus also of the gyro-screened potential $\phi_i$.

    In this model \textsc{flr} effects are assumed to be negligible for electrons (due to their small mass) but significant for ions. They are introduced through the gyro-average operators $\Gamma_0$ and $\Gamma_1$, which are expressed using the modified Bessel function $I_0$ in $k$-space: 
    \begin{align}\label{flr-operators}
        \Gamma_0 \equiv e^{- \tau k^2} I_0(\tau k^2)
        \quad \text{and} \quad
        \Gamma_1 \equiv e^{-\tau k^2/2}
    \end{align}
    with the ion to electron temperature ratio $\tau \equiv T_i/T_e$. The $\Gamma_0$ operator is implemented with the $\cosh$ approximated form presented in Ref. \citenum{olivares2018simple}. The classical drift-fluid modified Hasegawa-Wakatani model equations are recovered in the $\tau \rightarrow 0$ limit.

    All quantities above are dimensionless due to standard drift normalizations.  
    Spatial derivatives are normalized with the drift-scale $\partial_j \equiv \rho_0 \partial_j$ for $j \in (x,y)$, whereas derivatives in time are normalized with $\partial_t \equiv L_\perp/c_0 \partial_t$. The densities are normalized with a constant, static background density $n_0$ via $N_s \equiv N_s / n_0$ for both species $s \in (i,e)$ and the potential is normalized with $\phi \equiv \frac{e \phi}{T_e}$.

    \subsection{Code and Simulation parameters}
        The equations given above were solved with the gpu-parallelized \textsc{ghwc} code, which was forked from the open-source code \textsc{ghw}\cite{kendl2025ghw}. 
        
        The time-stepping is done with Karniadakis' Adams–Bashford scheme\cite{karniadakis1991highorder}, the Poisson-brackets are solved with Arakawa's scheme\cite{arakawa1997computational} and all other spatial derivatives are solved with fourth-order centered difference schemes. \textsc{flr} operators and the polarization equation are solved with spectral methods using the cu\textsc{FFT} library. More details on the solvers and on the algorithm can be found in Ref. \citenum{kendl2025ghw}, since \textsc{ghwc} uses similar methods and solvers as \textsc{ghw}. However, a publication of the new \textsc{ghwc} code as an update of \textsc{ghw} is intended in the near future.
                
        The scale ratio, which primarily affects the time-scale, was set to $\delta = 0.015$, while the background density gradient term is set to $g_n \equiv \frac{1}{\delta} \partial_x N_0 = -1$ assuming a constant, static, negative background gradient. Note that the behavior of the model with respect to the turbulence to zonal flow interaction can still be fully controlled\cite{numata2007bifurcation, guillon2025phase} through the parallel coupling parameter $C$.

        The initial conditions were set as a Gaussian perturbation  $N_s(\mathbf{x},0) = 0.1 \exp{(-(\mathbf{x} - \mathbf{x}_0)^2/16)}$ for both species with $\mathbf{x} = (x,y)$ and $\mathbf{x}_0 = (56,56)$. For ensembles of simulations an additional perturbation of $0.01$ was added at a random location to investigate sensitive dependence on the initial conditions.
        
        We test if the size of the initial perturbation influences the final size of the zonal flow modes by also testing it with different initial conditions that randomly initialize a "turbulent bath" with many different modes. This seems to hold in the range of the parallel coupling parameter $0.4 \geq C \geq 2$ investigated in this publication, where the system undergoes long ($\Delta t > 100 L_\perp/c_0$) turbulent phases before zonal flows emerge. However in the highly adiabatic regime $C \gg 1$ the size of the initial perturbation might influence the radial modes.
        
        Periodic boundary conditions were used in both directions.

        If not mentioned differently a square simulation domain of $L_x = 128 \rho_0$ together with a resolution of $n_\rho = 4$ nodes per length-unit $\rho_0$ were used. However, more details on simulation-domain and resolution are presented in section \ref{sec:domain_resolution}.

\section{zonal flow theory}\label{sec:zonal flow-theory}
    The zonal flow evolution equations are derived from the \textsc{mghw} model in this section. The gyroaveraging operators $\Gamma_0$ and $\Gamma_1$ from equations \eqref{flr-operators} are defined in Fourier space and are used in this form in the numerical algorithm. 
    
    However the formal analysis below is in configuration space therefore uses the second order Padé approximated forms of the operators
    \begin{align}\label{pade}
        \Gamma_0 \approx \frac{1}{1-\tau \nabla^2} \quad \text{and} \quad \Gamma_1 \approx \frac{1}{1-\frac{\tau}{2}\nabla^2}.
    \end{align} 
    In that case the approximations
    \begin{align}\label{pade-trick}
        \frac{\Gamma_0 -1}{\tau} \approx \Gamma_0 \nabla^2
    \end{align}
    also holds to second order.
    \subsection{Zonal momentum evolution}   

    Applying a partial time-derivative on equation \eqref{pol} and using the approximations \eqref{pade} and \eqref{pade-trick} yields the generalized vorticity equation 
    \begin{align}\label{generalized_vorticity}
        \partial_t \frac{\Gamma_0 - 1}{\tau_i} \phi = \partial_t \widetilde{N}_e - \partial_t \Gamma_1 \widetilde{N}_i \nonumber \\
        \Rightarrow
        \partial_t \nabla_\perp^2 \phi = \Gamma_0^{-1} \partial_t \widetilde{N}_e - \Gamma_1 \Gamma_0^{-1} \partial_t \widetilde{N}_i.
    \end{align}
    Applying the zonal average and a radial integral on equation \eqref{generalized_vorticity} then yields the equation for the time-evolution of the zonal flows momentum. 
    Note that $\phi$ and $\phi_i$ act as stream functions for the electron and ion gyrocenter-densities, respectively.
    Then
    \begin{align}\label{zf_momentum_warm_0}
         \partial_t \langle v_y \rangle_y =
        \int dx \langle  \Gamma_0^{-1} \partial_t N_e  \rangle_y
        -  \int dx \langle  \Gamma_1 \Gamma_0^{-1} \partial_t N_i \rangle_y.
    \end{align}
    Inserting equation \eqref{cont_ne} and \eqref{cont_ni} into equation \eqref{zf_momentum_warm_0} and using Reynolds decomposition $f(x,y) = \langle f(x) \rangle_y + \hat{f}(x,y)$ for all scalar fields $f$ involved, yields the zonal flow momentum evolution equation or, in other words, the zonal flow momentum conservation equation
    \begin{align}\label{zf_momentum_warm}
        \partial_t \langle v_y \rangle_y = 
       \frac{\delta^{-1}}{\Gamma_0}
        ( 
            \langle \hat N_e \partial_y \hat \phi \rangle_y
            - \Gamma_1 \langle \hat N_i \partial_y \hat \phi_i \rangle_y
        )
        - \nu \partial_x^4 \langle v_y \rangle_y.
    \end{align}

     In the derivation, the parallel coupling term as well as the gradient-drive term vanish in the zonal average when applying partial integration. Note that the gyro-operators (in this case) do not have any spatially variable coefficients. Also note that the integration domains are constant and the boundary values for the integration with respect to $x$ vanish with periodic boundary conditions. Therefore, the gyro-operators can be pulled out of the integrals.
    
    Analyzing the transfer terms in the right hand side of equation \eqref{zf_momentum_warm} by taking the zero-\textsc{flr}, cold-ion limit recovers the radial gradient of the well known Reynolds stress \cite{diamond1991theory, diamond2005zonal, scott2005energeticszf, itoh2006zfphysics, gurcan2015zonal, held2018non, Scott1} for the first term
    \begin{align}\label{zf_momentum_cold}
         \partial_t \langle v_y \rangle_y \overset{\tau = 0}{\longrightarrow} 
        \delta^{-1} \partial_x \langle \hat{v}_x \hat{v}_y \rangle_y
        - \nu \partial_x^4 \langle v_y \rangle_y.
    \end{align}
    
    The hyperviscosity contribution to the zonal flow momentum evolution in equation \eqref{zf_momentum_warm} remains untouched by \textsc{flr} operators.

    In equation \eqref{zf_momentum_cold} it can be easily seen, that the total zonal flow momentum (even with the hyperviscosity contribution) remains conserved, since an integral over the radial $x$-domain would vanish in periodic boundary conditions. This means that zonal flow momentum is transferred locally, but conserved globally.
    
    \subsection{Zonal energy evolution}
        The zonal energy is defined as 
        \begin{align}\label{zonal-energy}
            E_{zf} \equiv \frac{1}{2} \left( \partial_x \langle \phi_i \rangle_y \right)^2.
        \end{align}  
        Then the zonal energy evolution equation can be derived by 1) applying the flux-surface average on the polarization equation, 2) taking the partial time derivative on it and 3) multiplying it with $\langle \phi \rangle_y$. The final step is 4) inserting continuity equations \eqref{cont_ne} and \eqref{cont_ni} and applying the integral of the 2D volume $V$ on the result. This yields the zonal energy evolution equation
        \begin{align}\label{zf_eng_evolution}
            \partial_t &\int dV E_{zf} = \\ \nonumber
            \int dV \frac{1}{\delta} \bigg( &\big( 
                \langle \phi \rangle_y \langle \hat N_e \partial_y \hat \phi \rangle_y -
                \langle \phi_i \rangle_y \langle \hat N_i \partial_y \hat \phi_i \rangle_y \big) \\ \nonumber
                - \nu &\big( 
                    \langle \phi_i \rangle_y \partial_x^4 \langle N_i \rangle_y -
                    \langle \phi \rangle_y \partial_x^4 \langle N_e \rangle_y 
                    \big) \bigg).
        \end{align}
        Applying the $\tau \rightarrow 0$ limit on equation \eqref{zf_eng_evolution} gives
        \begin{align}\label{zf_eng_cold}
            \overset{\tau = 0}{\rightarrow}
             \int dV  \bigg(
            \big( \frac{1}{\delta} \langle \hat{v}_x \hat{v}_y \rangle_y \partial_x \langle v_y \rangle \big)
            - \big( \nu \langle v_y \rangle \partial_x^4 \langle v_y \rangle \big) \bigg)
            \nonumber \\
            = \int dV  \bigg(
            \big( \frac{1}{\delta} \langle \hat{v}_x \hat{v}_y \rangle_y \partial_x \langle v_y \rangle \big)
            - \nu \big( \partial_x^2 \langle v_y \rangle \big)^2 \bigg).
        \end{align}
        which also recovers the expected term with the radial derivative of the mean flow multiplied with Reynolds stress in Ref. \citenum{Scott1}. The last step uses partial integration with periodic boundary conditions.

        Note that no total conservation of energy can be observed in equation \eqref{zf_eng_cold}. On the contrary, since $\nu > 0$ and $(\partial_x \langle v_y \rangle  )^2 > 0$ it can be concluded, that the negative sign of the hyperviscosity contribution suppresses zonal flows over time, if no Reynolds stress is present as a source of zonal flow energy.

\section{Why do zonal flows merge?}\label{sec:why_zf_merge}

    In Ref. \citenum{grander2024hysteresis}, it was shown that a significant change of zonal energy and transport can happen suddenly even after the system has been seemingly stabilized in a nonlinear equilibrium. This sudden rise of zonal energy coincides with a decrease of the radial zonal flow modes. Two streams in one direction seem to \emph{'merge'} and extinguish the stream in the other direction, which has been between them before. A similar observation was reported in Ref. \citenum{ivanov2020zonally} for two-dimensional ITG turbulence.

\begin{figure}
        \centering
        \includegraphics[trim=0cm 0.5cm 0cm 0.1cm,clip=true,width=\linewidth]{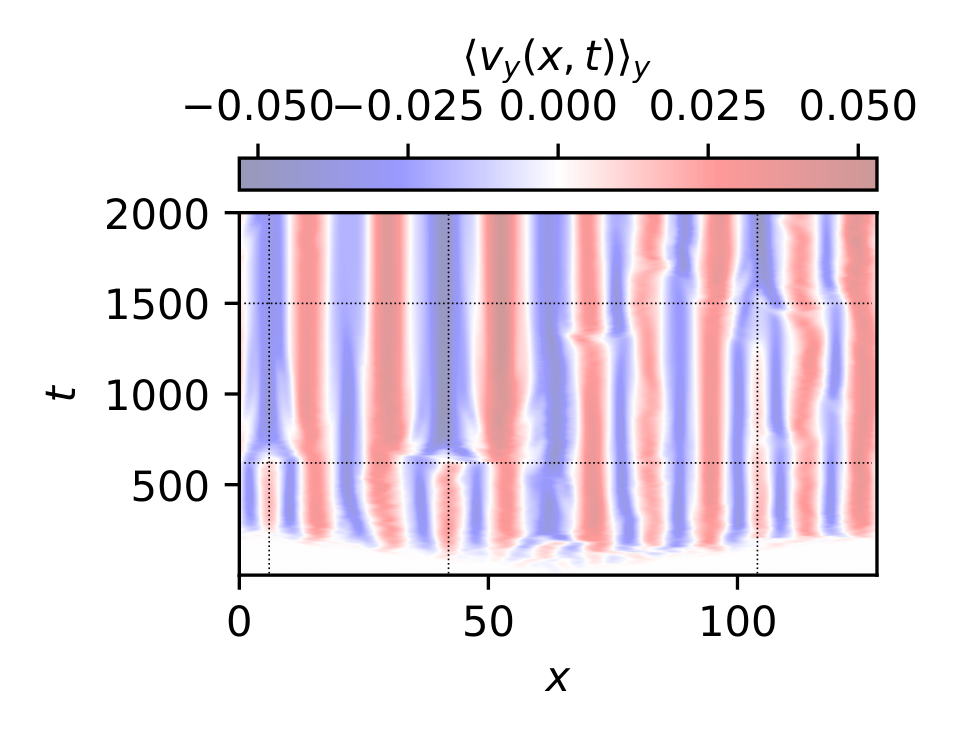}
        \caption{Zonal velocity profile for $\tau=0$ and $C=1.7$. After the development of a seemingly stable zonal flow profile up to  $ \sim 500 L_\perp/c_0$ two streams vanish at $x_1 \approx 6\rho_0$ and $x_2 \approx 42 \rho_0$  (left and middle vertical lines) at $t_1\approx 620$ (lower horizontal line). Another flow vanishes at $x_3 \approx 104 \rho_0$ and $t_2 \approx 1500$ (upper horizontal and right hand vertical line).}
        \label{fig:what_is_merging}
    \end{figure}

\begin{figure*}
        \includegraphics[width=\linewidth]{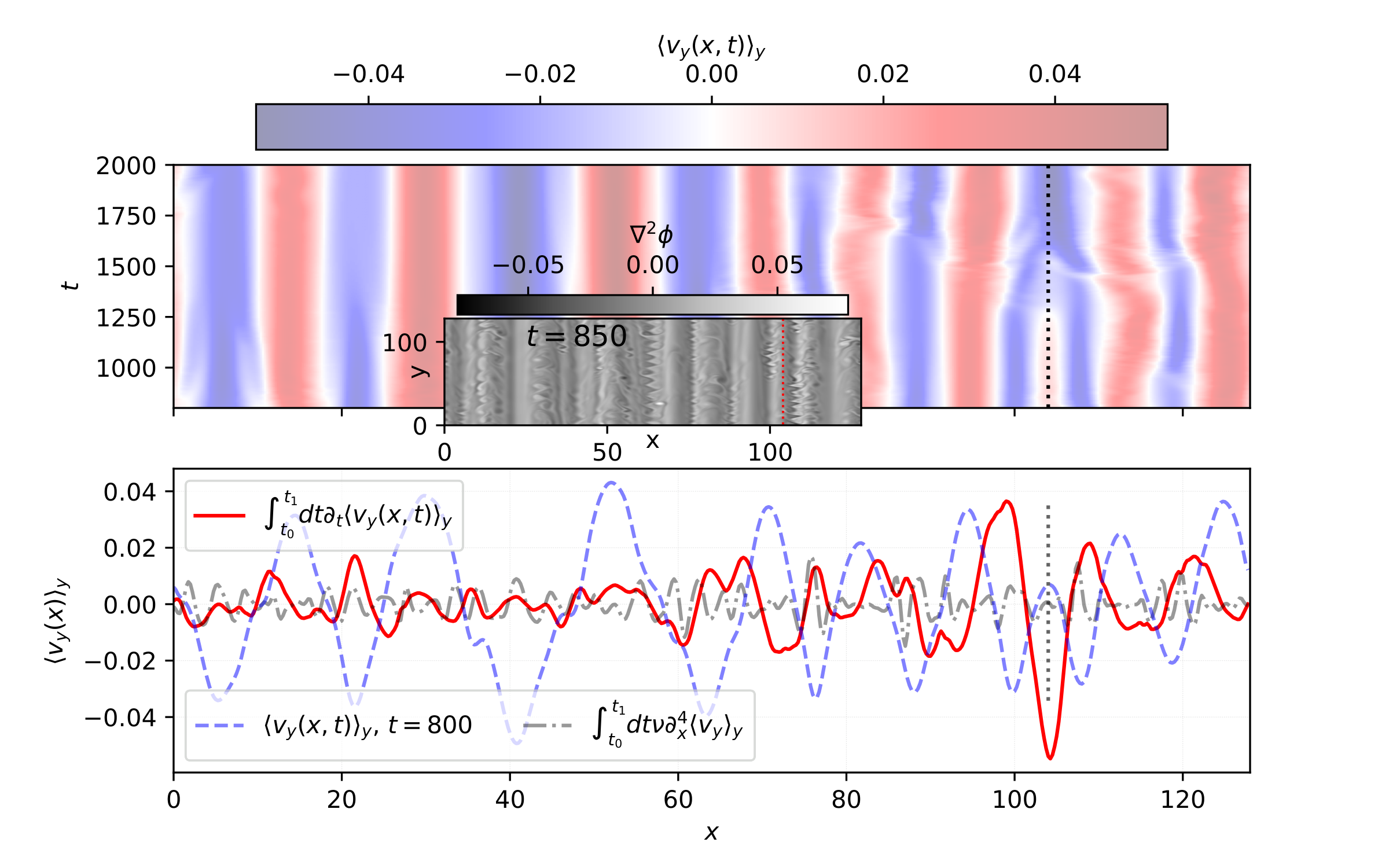}
        \caption{\label{fig:why_zf_merge}. 
        Merging of a zonal flow for the same simulation as presented in Figure \ref{fig:what_is_merging} with parameters $\tau_i=0, C=1.7$. The top panel depicts the zonal flow velocity profile $\langle v_y(x,t) \rangle_y$ from $t_0=800$ up to $t_1=2000L_\perp/c_0$. The dotted vertical lines at $x=104\rho_0$ show the location of the merger in the top and the bottom panel.
        The inset shows the vorticity profile at $t=850 L_\perp/c_0$. Non-zonal vorticity structures are present in the vicinity of the merger (red, vertical, dotted line) and also in other radial locations.
        The bottom panel shows the total cold-ion zonal flow momentum drive (the right-hand side of equation \eqref{zf_momentum_cold}) integrated over the whole depicted time-frame (red, solid line) and the hyperviscous contribution to the total drive (black, dash-dotted line). 
        Also the zonal flow profile at $t=800 L_\perp/c_0$ (blue, dashed line) is shown in the bottom panel. 
        At the radial location of the merger $x = 104\rho_0$ the momentum drive is destructively interfering with the zonal flow momentum. The momentum drive is also interfering with a downwards flow at $x \approx 20\rho_0$. However, in this case it only leads to a weakening of the flow and not to a merger.
        Note that the hyperviscous contribution tends to be small but not negligible compared to the radial gradient of the Reynolds stress in equation \eqref{zf_momentum_cold}.
        }
    \end{figure*}

    The color bars in Figures \ref{fig:what_is_merging} and \ref{fig:why_zf_merge} (top panel) both show the direction and amplitude of zonal velocity profiles flows in $y$-direction $\langle v_y(x,t) \rangle_y$. Red streams go towards the positive $y-$direction (upwards), blue streams go towards the negative $y-$direction (downwards). Note that the positive upwards flow direction corresponds to the electron diamagnetic drift direction, which is also the flow direction of the basic drift-wave mechanism\cite{kendl2008two}. 

    In Figure \ref{fig:what_is_merging} three merger-events are presented in a simulation for $\tau = 0$ and $C = 1.7$. Three flows in positive $y$-direction (red) suddenly vanish or, in other words, the neighboring flows in negative $y$-direction (blue) merge at these points. 

    We noted in many simulations that if a merger happens, then in virtually every instance a (red) upward stream is vanquished by the merging of two (blue) downward streams. This break of symmetry might be connected to an asymmetry in the zonal flow profiles themselves. In Ref. \citenum{kammel2011analysis} it was reported in the context of zonal flows in the Hasegawa-Wakatani system, that upwards streams tend to be less sharp than downward streams. This can also be observed when estimating the curvatures of the zonal flow profile (blue, dashed line in the bottom panel of Figure \ref{fig:why_zf_merge}) by numerically calculating the second derivatives at the peaks (upward flows) and the troughs (downward flows). The mean curvature for the troughs ($\approx 0.015$) is almost twice as large than the mean curvature for the peaks ($\approx 0.008$). According to equation \eqref{zf_eng_cold} the nonlinear drive of zonal energy is proportional to the radial gradient of the poloidal flow $\partial_x \langle v_y \rangle_y$. Therefore the total nonlinear zonal energy drive could be smaller at the broader tops of the upwards flows compared to the sharper troughs of the downwards flows. This interrelation might be part of the explanation for the apparent asymmetry in what flows are merging. 
   Another kind of symmetry break in zonal flows that could influence this particular behavior by locking certain turbulent modes either in upwards or in downwards flows was recently reported by Ref. \citenum{sasaki2026bifurcation}.
    
\begin{figure}
        \centering
        \includegraphics[trim = {0 13 0 10}, clip, 
        width=\linewidth]{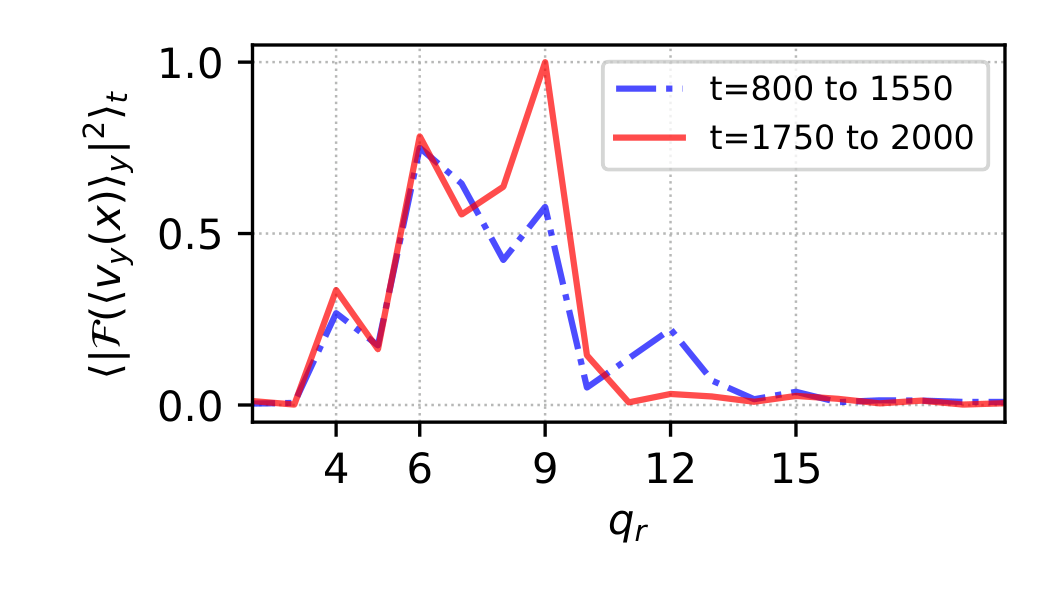}
        \caption{The time-averaged, normalized spectrum of the zonal flow profile before and after the merger for the simulation shown in Figure \ref{fig:why_zf_merge} ($\tau_i=0, C=1.7$). 
        The blue dash-dotted line is the spectrum averaged over the time before the merger $t=800$ to $t=1550L_\perp/c_0$. 
        The red, solid line is the spectrum after the merger $t=1750$ to $t=2000L_\perp/c_0$. 
        One peak at $q_r = 12$ exists in the spectrum only before the merger. The peak at $q_r = 9$ appears amplified after the merger. 
        The dominant radial mode-numbers are $q_r^d = 6$ initially and $q_r^d = 9$ after the merger.
        For higher wave-numbers the spectra are virtually indistinguishable.}
        \label{fig:merge_spectral}
    \end{figure}

    A closer look at the \emph{'merger'} happening at $(x,t) \approx (104\rho_0, 1500L_\perp/c_0)$ is presented in Figures \ref{fig:why_zf_merge} and \ref{fig:merge_spectral}. 
    
    Merging events appear to happen due to the same reason, why zonal flows emerge in the first place: momentum is locally transferred through nonlinear interactions given by the radial gradient of the Reynolds stress and a smaller (but not negligible) hyperviscosity contribution according to equation \eqref{zf_momentum_warm}. 
    
    This can be seen in the bottom panel of Figure \ref{fig:why_zf_merge}. The time integral over the right hand side of the cold-ion zonal flow momentum drive acts as a local transfer-term of zonal flow momentum. A significant minimum is exactly aligned with a zonal flow velocity peak at the location of the merger ($x = 104\rho_0$).  
    This indicates that momentum is decreased through the nonlinear interaction at this location throughout the presented simulation-time until the upward flow vanishes at the given radial location.
    
    Another observation emphasizing the nonlinear nature of this phenomenon is the fact, that visually distinguishable non-zonal vorticity structures always seem to be present at the location of merging events. The inset in Figure \ref{fig:why_zf_merge} shows the vorticity, which is smooth in some parts and structured in other parts. We propose that non-zonal structures with small, radial expansion in the vorticity profile can serve as a rough proxy for the nonlinear zonal flow momentum drive, because they are a necessary condition for a finite radial gradient of the Reynolds stress. 
    
    Figure \ref{fig:merge_spectral} shows time-averaged radial spectra of the zonal flow profile $\langle|\mathcal{F}_x(\langle v_y(x,t) \rangle_y |^2 \rangle_t$ for the simulation data presented in Figure \ref{fig:why_zf_merge} with the radial mode numbers $q_r(k_x) \equiv k_x  L_x$ as the horizontal axis. The spectra were calculated in python using the \texttt{numpy.fft} library.
    As already observed in Ref. \citenum{grander2024hysteresis} (there in Fig. 7) the visually dominant zonal flow modes - the number of red or blue stripes in the velocity profiles - are not necessarily the dominant ones in the spectrum.
    
    The merger seems to shift the spectrum towards the lower modes, since the peak at $q_r = 12$ vanishes subsequently and the $q_r = 9$ mode is amplified almost by a factor of two. The visually dominant modes in this simulation are $q_r=9$ before and $q_r=8$ after the merger (see Figure \ref{fig:why_zf_merge}). However, the actual dominant radial mode shifts from $q_r=6$ to $q_r=9$ in the spectrum, because of the amplification of the peak at $q_r=9$, even though the merger did not effect the amplitude of the $q_r=6$ mode significantly.

     We conclude this section with the proposition that, both the merging of zonal flows and the generation of zonal flows from turbulence in the first place, are (mostly) nonlinear phenomena with the same cause: the radial gradient of the Reynolds stress and the contribution of hyperviscosity given in equations \eqref{zf_momentum_warm} and \eqref{zf_momentum_cold} respectively. In general we also observe that mergers shift the radial spectrum towards lower mode-numbers (see also Figure \ref{fig:zf_mode_picking} in the subsection below).
    
    Since both the emergence and the evolution of zonal flows appear to be nonlinear processes one might ask the following two questions: 1) Which zonal flow patterns emerge in the first place? 2) Is there an equilibrium state at all, or can mergers always happen in a driven system like the Hasegawa-Wakatani model? 

    \subsection{Which zonal flow patterns emerge in the first place?}
        
        We attempt to answer to the first question with the following experiment: 100 simulations for $C=0.4$ and $\tau=0$ are conducted up to $t=300L_\perp/c_0$ and another set of 100 simulations up to $t=3000L_\perp/c_0$ respectively with exactly the same parameters. The densities in each simulation are initialized with a Gaussian blob as described in section \ref{sec:model_and_code} with an additional perturbation of 1\% of the blob's amplitude at a random location.

        We describe the zonal flow pattern of one simulation at a given time $t$ with the dominant radial-mode number $q_r^d(t) \equiv \max_{q_r} |\mathcal{F}_x(\langle v_y(x,t) \rangle_y |^2$ because it approximates the number of peaks or streams in the zonal flow profile for a given radial domain length $L_x$. The dominant radial wavenumber can be simply calculated with $\lambda_r^d = L_x / q_r^d$.
        
        Of course, as already discussed in the section above in Figure \ref{fig:merge_spectral}, other modes $q_r$ contribute to the zonal flow profile as well. Also the visually discernible number of streams does not necessarily coincide with the spectrally measured dominant mode number. Nevertheless $q_r^d(t)$ can serve as a simple and illuminating metric to describe the zonal flow profile at a given time.
        \begin{figure}
            \centering
            \includegraphics[trim = {0 13 0 10}, clip, width=\linewidth]{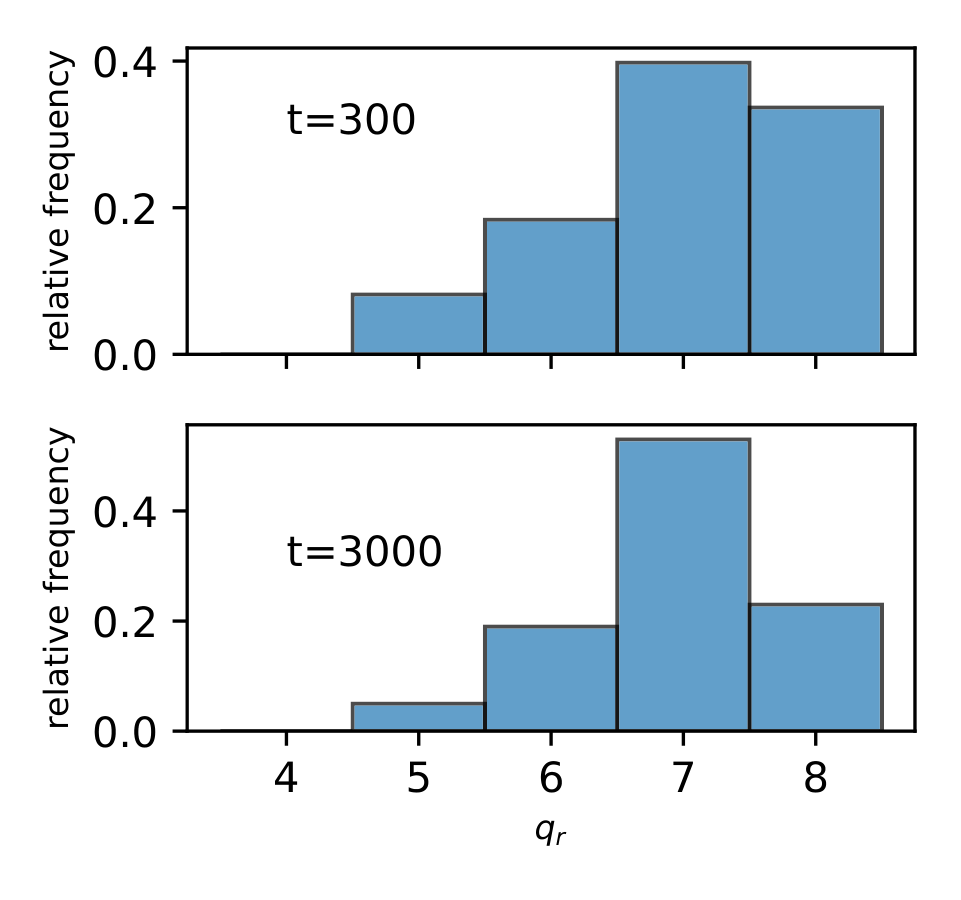}
            \caption{The distribution of the dominant radial mode numbers $q_r^d$ after $t=300 L_\perp/c_0$ (top panel) and $t=3000 L_\perp/c_0$ for 100 simulations respectively with $\tau=0$ and $C=0.4$. Each of the simulations was initialized with a random perturbation added to a Gaussian blob. Top panel: About 40 \% of the simulations show a dominant radial mode-number of 7, which corresponds to a wavelength of $\lambda_r \approx 18 \rho_0$. Bottom panel: Approximately 50\% of the simulations have the dominant radial mode-number of 7. In both cases other dominant mode-numbers are also present for $5 \leq q_r \leq 8$.}
            \label{fig:zf_mode_picking}
        \end{figure}
        In Figure \ref{fig:zf_mode_picking} (top panel) we show that there is a distribution of the dominant radial mode numbers $q_r^d$ when zonal flows emerge out of turbulence at a simulation time of $t=300 L_\perp/c_0$ (Figure \ref{fig:zf_mode_picking}, top panel). The variation in the dominant radial mode-number occurs due to small changes in the initial conditions. This indicates that there are chaotic, nonlinear processes at play in the choice of the initial zonal flow patterns, as it has been observed before for collisionless gyro-kinetic ion-temperature-gradient simulations of zonal flows\cite{dif2008influence}. Note that $300 L_\perp/c_0$ is close to the time when zonal flows appear in the simulation out of turbulence for the chosen parameters (see Figure \ref{fig:what_is_merging}). 
       
        \subsection{Effects of resolution and domain-size on zonal flow patterns}\label{sec:domain_resolution}

        A prior convergence study has been conducted in Ref. \citenum{mastersthesis} with the result that a resolution of $n_\rho = 4$ and a simulation domain of $L_x = 128 \rho_0$ are sufficient values for a parameter range of $0.01 \leq C \leq 1$ and $0 \leq \tau \leq 1$. In that study, time-averages of values in nonlinear equilibrium of single simulations were compared for different parameters. Also the Padé approximations for the \textsc{flr} operators were used in the former study. It was noted in Ref. \citenum{mastersthesis} that the results were partially inconclusive for high values of the parallel coupling parameter $C$.
        
       In this section, we reproduce the results of this study with an emphasis on zonal flow patterns in order to verify, that the simulation parameters make sense for the experiments presented in this publication. Also, we use the \textsc{flr} operators in their original forms as presented in equations \eqref{flr-operators}. 
       Ensembles of simulations with small variations in the initial conditions are used to produce distributions of the dominant radial wavenumbers $q_r^d$ as presented in Figure \ref{fig:zf_mode_picking}. This takes into account the chaotic dependence of zonal flow patterns on small (random) variations of the initial conditions. The mean values and standard deviations of this distributions can then be compared for different sets of parameters.

\begin{figure}
            \begin{subfigure}{0.4\textwidth}
                \includegraphics[width=\textwidth]{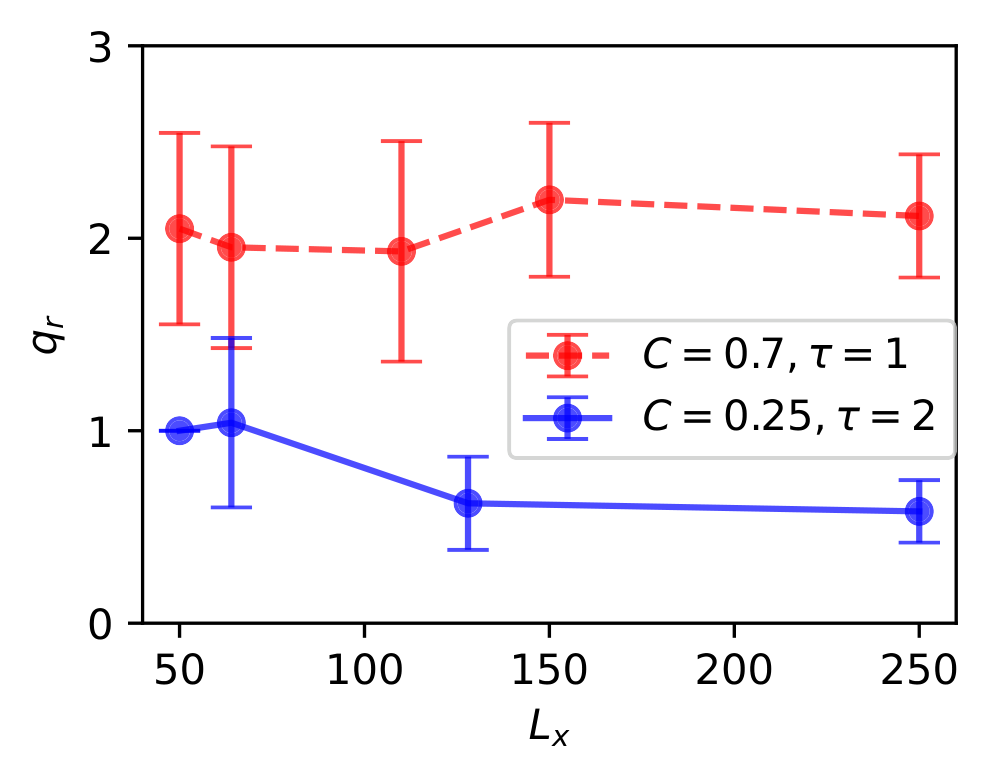}
                \caption{Distributions of the dominant radial mode-number $q_r^d$ for different square box-sizes $L_x$ at $t=3000L_\perp/c_0$. All simulations have the same resolution $n_\rho = 4$ with the hyperviscosity factor $\nu = 5 \times 10^{-4}$. The mode-numbers were scaled to the smallest box $L_x = 50\rho_0$.}
                \label{subfig:domain_check}
            \end{subfigure}
            \hfill
            \begin{subfigure}{0.4\textwidth}
                \includegraphics[width=\textwidth]{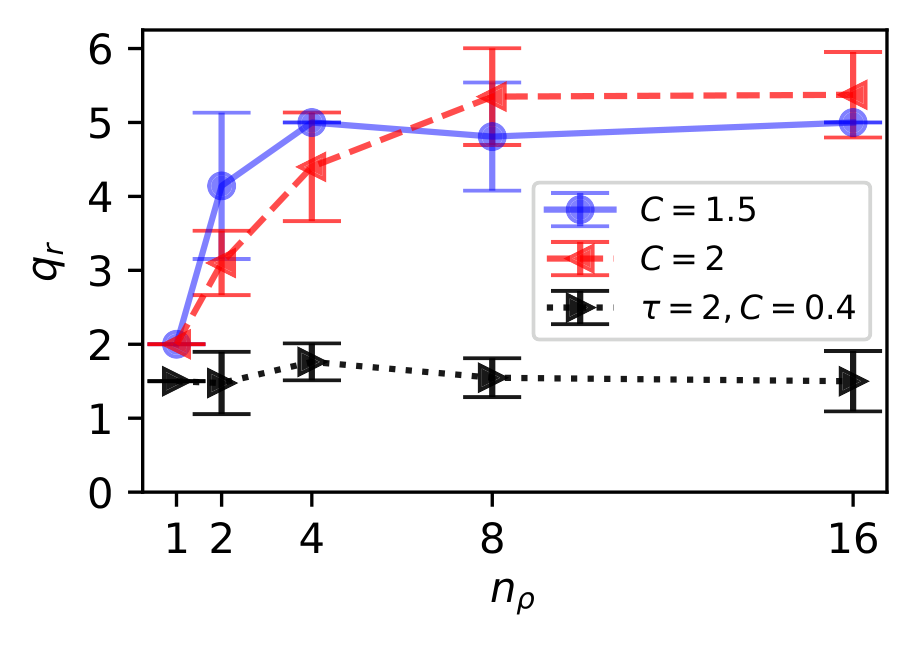}
                \caption{Distributions of $q_r^d$ at $t=500L_\perp/c_0$ for different resolutions $n_\rho$ with a square box-size $L_x = 64\rho_0$ for $\tau = 0$ and $L_x = 128 \rho_0$ for $\tau = 2$ (black, dotted, values scaled to the smaller box). The hyperviscosity parameter is scaled with $\nu \approx 0.125 (1/n_\rho)^4$.}
                \label{subfig:resolution_check}
            \end{subfigure}                  
            \caption{Mean values of the dominant radial mode numbers $q_r^d$ for different values of domain-size (top) and resolution (bottom) respectively of at least twenty simulations for each value. The error-bars are the standard deviations.}
            \label{fig:boxsize_resolution}
        \end{figure}

        In Figure \ref{fig:boxsize_resolution} we present convergence of resolution and simulation domain using this method.
        
        Figure \ref{subfig:domain_check} addresses the question whether the simulation domain is large enough to cover the dominant radial modes. We observe that for low values of the parallel coupling $C \approx 0.25$ together with the high value of $\tau = 2$ convergence appears to be reached (within error-bars) at $L_x \geq 100\rho_0$. For larger values $C \geq 0.7$ in combination with smaller values of $\tau < 2$ it seems to be reached only at approximately $L_x \geq 100 \rho_0$.
        
        In general, we observed that structures tend to be larger for \textit{small} values of $C$ compared to intermediate values\cite{Scott1}. Note that this trend does not continue into the highly adiabatic regime $C \gg 1$, where the stronger inverse cascade\cite{camargo1995resistive} then produces larger structures for higher values of $C$. Also \textit{large} values of $\tau$ tend to result in larger structures. Since $C=0.25$ is the smallest value where zonal flows appear for $\tau=2$, we can conclude from Figure \ref{subfig:domain_check}, that $L_x = 128 \rho_0$ is sufficient for all the parameters investigated here and $L_x = 64 \rho_0$ is sufficient for $C \geq 0.7$.
        
       In Figure \ref{subfig:resolution_check} we present the convergence with respect to resolution. However also finite values of $\tau$ can cause small structures in the vorticity\cite{kendl2018vortex} and thus also may demand a higher resolution.
       A necessary and sufficient resolution of $n_\rho = 4$ for parallel coupling $C = 1.5, \tau = 0$ matches the results from Ref.\citenum{mastersthesis}. We conclude, that the result also holds for smaller values $C\leq 1.5$. In particular this resolution appears to be sufficient for $C=0.4$ regardless of the value of $\tau$.      
       In the case of $C = 2, \tau = 0$ the resolution of $n_\rho = 4$ is barely within the error-bars (standard deviation) of the higher resolutions. Using that resolution is perhaps acceptable, but a higher resolution might be recommendable.

       For the highly adiabatic regime $C \gg 1$ even a higher resolution ($n_\rho = 16$ for $C = 10$) appears to be necessary (data not shown here).
    
\section{Does an equilibrium state exist?}\label{sec:equilibrium-state}

        In this section we discuss the possibility of an actual equilibrium state for simulations, where zonal flows are present. 

        We observed, that the merging of zonal flows tends to be a one-way process (towards lower mode-numbers). On the contrary, it was observed for the ITG case\cite{ivanov2020zonally}, that zonal flows can also split up and thus in principle reverse the process of merging. In a parameter range of $0.1 \leq C \leq 2$ and $0 \leq \tau \leq 2$ we were unable to verify the existence of this reverse process of splitting up of zonal flows. 
        
        Only in the highly adiabatic regime $C \gg 1$ we observed fast radial propagation of zonal flow momentum as an addition to the zonal flow dynamics described above. However, that effect appeared to be less pronounced or even non-existent for higher resolutions (and lower hyperviscosity parameters). It might thus be caused mainly by viscosity and therefore not be that relevant for the simulation of a hot plasma. Also for a very high value of $\tau = 4$ one simulation with a split up of zonal flows was discovered (data not shown here). An investigation of those more extreme parameter ranges is computationally expensive and postponed to future work.

        If the merging of zonal flows is in fact a one-way process for this particular model and the parameter range discussed here, the following question arises: Will the flows keep merging over time and end up filling the whole simulation domain with one flow $q_r^d = 1$? Or does a stable stable zonal flow pattern exist, where no mergers will happen anymore once it has been reached?

        In Figure \ref{fig:zf_mode_picking} we present observations regarding the distribution of dominant radial mode-numbers at the onset of the zonal flows as well as in later times.
        
        The distribution of dominant mode-numbers $q_r^d$ at $t=3000 L_\perp/c_0$ (Figure \ref{fig:zf_mode_picking}, bottom panel) have a more pronounced peak around the mode-number $q_r^d = 7$ than the one for $t=300$ (top panel). 
        The absolute number of simulations that show modes lower than $q_r^d < 7$ increases in the longer simulations compared to the shorter ones (from 24 to 26) whereas the absolute number of simulations with $q_r^d > 7$ is decreased (from 35 to 26). Thus the system appears to favor a certain dominant mode-number, which depends on the simulation parameters but not so much on the initial conditions. In this specific case $(C=0.4, \tau = 0, L_x = 128, n_\rho=4)$ we observe, that in both sets of 100 simulations $q_r^d=7$ is the dominant mode-number in 40\% for $t=300$ and then 50\% for $t=3000$ of the simulations respectively. 
        
        However, while there seem to be favored zonal flow modes for a given set of parameters, it is still not certain \emph{a priori} which $q_r(t)$ will appear in one single simulation at any given time $t$. Furthermore, we were able to detect mergers as late as $t \approx 1.5 \times 10^5 L_\perp/c_0$ and with $q_r^d \approx 3$ for $t \in [1.5\times 10^5, 5\times 10^5]$ ($\tau=0, C=0.4$, simulation data not shown here). On one hand this implies the possibility of a shifting of the favored $q_r^d$ towards smaller mode numbers for longer simulations. On the other hand we could not verify that simulation eventually achieve $q_r^d = 1$, where no more mergers are possible. Unfortunately large ensembles of long simulations ($\approx 10^5 L_\perp/c_0$) are computationally very expensive and for the moment outside of the scope of this publication.
        
       Thus it remains unclear empirically, if the state of any simulation involving zonal flows is ever absolutely stable. 
               
     Theoretically the equations \eqref{zf_momentum_warm} and \eqref{zf_momentum_cold} imply the existence of equilibrium zonal flows in the cold ion limit, when neglecting the hyperviscosity contribution.
     
     Note that in Ref. \citenum{guillon2025phase} the viscosity contribution only acts on the non-zonal components of the time-evolved quantities and therefore would not appear at all in the zonal flow momentum \eqref{zf_momentum_warm} and energy \eqref{zf_momentum_cold} equations respectively. In that case only the nonlinear term with the Reynolds stress appears as a contribution. Unfortunately we were not able to obtain numerically stable simulations in our code framework using this specific non-zonal viscosity term. Besides the trivial solution, any system of orthogonal periodic functions would effectively be stable, while any flow that has a finite overlap of $\hat{v}_x $ and $\hat{v}_y $ will be unstable. As the orthogonality relation must hold for the non-zonal velocity components, the corresponding potential and density distribution is non-trivial and does not occur naturally. Orthogonal configurations like 
    $
    \hat{v}_x =\cos{x}\cos{y}, \hat{v}_y =\sin{x}\sin{y}
    $
    are technically stable but, similarly to the co-advecting vortex problem, become unstable after some time due to numerical rounding errors \cite{kendl2018vortex}.
    Thus in theory only trivial situations with $\hat{v}_x=0$ or $\hat{v}_y=0$ enable analytically stable zonal flow configurations without hyperviscous contributions to the zonal velocity components. 
       
   One way to ensure this stability of zonal flows would be the suppression of all medium-scale ($\approx \rho_0$) non-zonal structures $\hat{v}_x, \hat{v}_y$ through (hyper)-viscous damping, which would not make sense from a modeling perspective.
   
    We conclude that once zonal flows play a role for quantities like the radial transport, they should be preferably established through ensembles of simulations with varying initial conditions and not through single simulations, since the initial zonal flow patterns are chosen chaotically and the existence of a final equilibrium state could not be established analytically or numerically with the parameters given in section \ref{sec:model_and_code}.

\section{Finite Larmor radius effects, hysteresis and phase transitions in zonal flows}\label{sec:flr-and-stuff} 

    \subsection{Finite Larmor radius effects on zonal flow patterns}
    
\begin{figure}
        \begin{subfigure}{0.4\textwidth}
            \includegraphics[width=\textwidth]{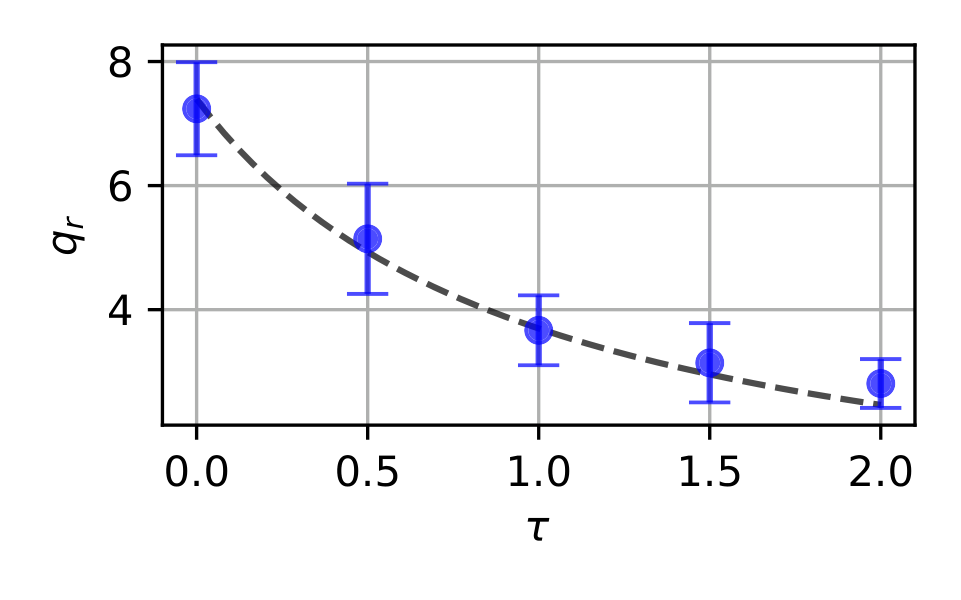}
            \caption{The dominant radial mode-numbers in dependence of the ion to electron temperature ratio $\tau$ for $C = 0.4$. The black dashed line $q_r^d(C=0.4,\tau) \equiv 7.4/(1+\tau)$ was fitted to the data points.}
            \label{subfig:tauvsqr}
        \end{subfigure}       
        \hfill
        \begin{subfigure}{0.4\textwidth}
            \includegraphics[width=\textwidth]{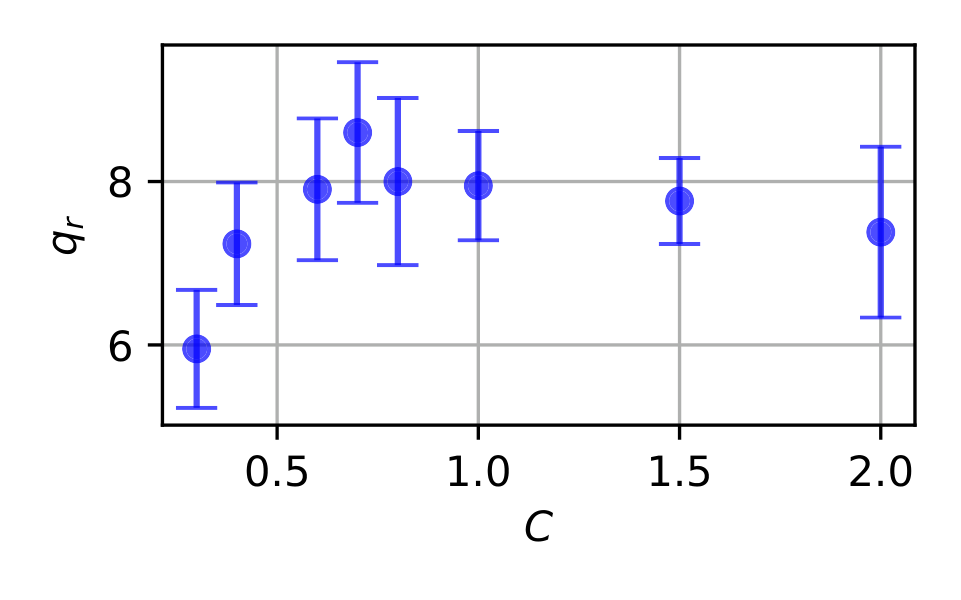}
            \caption{The dominant radial mode-numbers in dependence of the resistive coupling parameter $C$ for $\tau = 0$. The values appear to increase up to $C \approx 0.7$ with no further change for higher values of $C$.}
            \label{subfig:cvsqr}
        \end{subfigure}                
        \caption{The mean values of distributions of dominant radial mode numbers $q_r^d$ for different combinations of $(\tau, C)$ at $t=3000L_\perp/c_0$ for $21$ simulations each. The error-bars show the standard deviations. Larger values of $\tau$ result in consistently lower values of the dominant wave-numbers $q_r^d$.}
        \label{fig:dom_mode_scan}
    \end{figure}

    In this section we report the results of a parameter scan over values of the ion to electron temperature ratio $0 \leq \tau \leq 2$ and an intermediate range of the parallel coupling parameter $0.4 \leq C \leq 2$. Significant \textsc{flr} effects on the dominant radial mode number could be observed and are presented in Figure \ref{fig:dom_mode_scan}. 
    
    Visually, the zonal flows are wider and less smooth for finite $\tau$ in comparison with the cold-ion case. This is consistent with the fact that coherent structures in density tend to be larger and smeared out\cite{kendl2018vortex} for finite values of $\tau$. Empirically, a linear decrease of the dominant radial mode-number
    \begin{align}
        q_r^d(\tau) \approx \frac{q_r^d(\tau = 0)}{1 + \tau}
    \end{align}
    was observed and is presented for $C = 0.4$ in Figure \ref{subfig:tauvsqr}.

    The scan over the parallel coupling parameter did not yield a clear tendency for the whole range, which is in line with the previous result reported in Ref. \citenum{numata2007bifurcation} (there in Figure 7). However a linear increase might be discernible for small values of $C < 0.8$ before saturating or even decreasing then for larger values of $C$. This would roughly correspond to the spectra shown in Ref. \citenum{camargo1995resistive} and Ref. \citenum{Scott1} (there in Figures 5 and 4.15 respectively).

\subsection{Hysteresis and phase transitions in the drift-wave to zonal flow transition}

\begin{figure*}
    \includegraphics{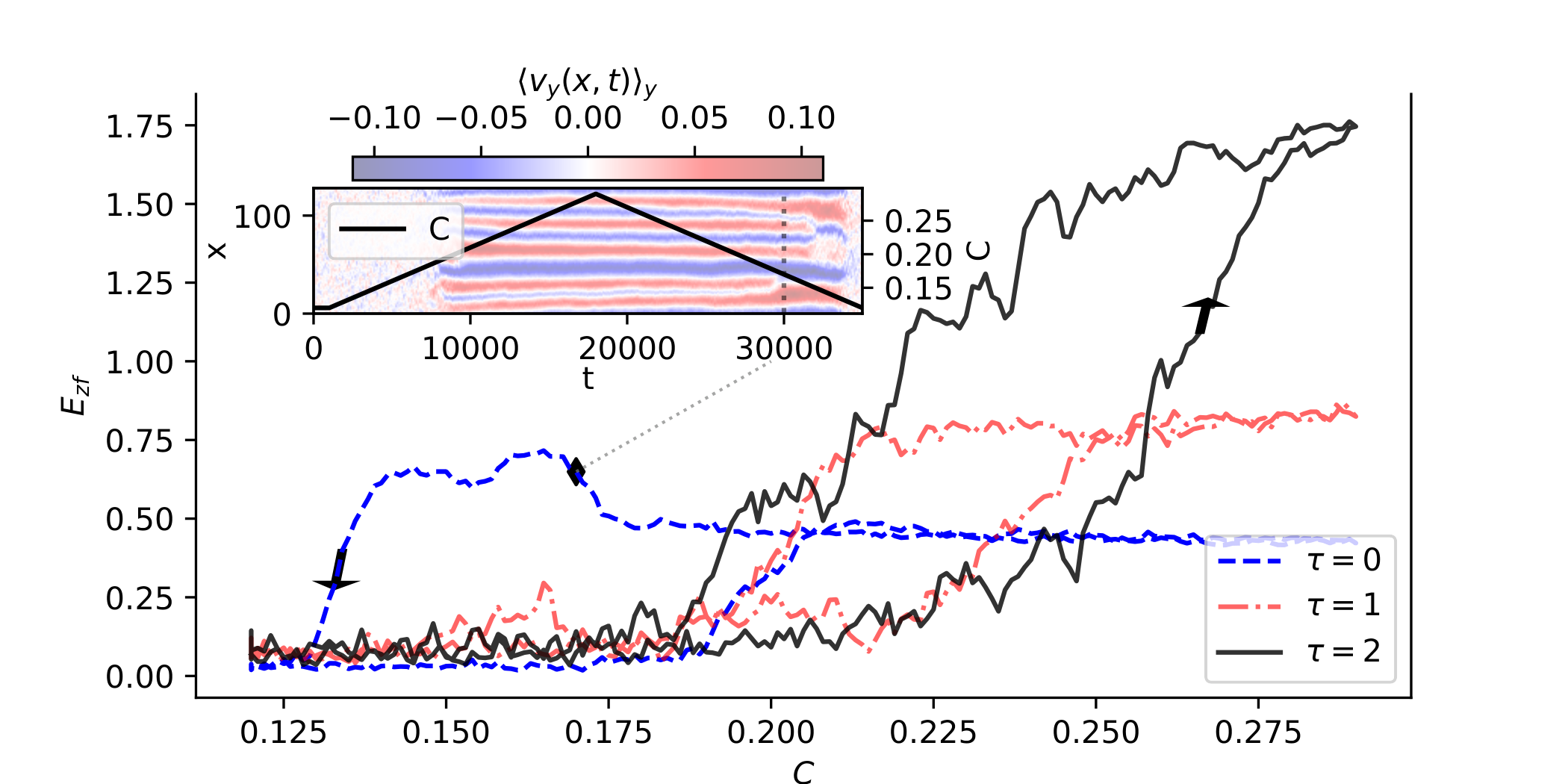}
     \caption{The zonal energy from equation \eqref{zonal-energy} vs. the resistive coupling parameter $C$ for different values of $\tau \in (0,1,2)$ respectively. For every simulation $C$ is ramped up and down the interval $C \in [0.12,0.29]$, with a slope of $ \Delta C/ \Delta t = -0.01/1000$ within the time-frame of $t \in [1000,30000]$. $C$ is constant in the initial phase.
     The arrows point in the positive time direction. 
     The inset shows the zonal flow velocity profile in time for the $\tau = 0$ case (blue, dashed line) with the radial position on the left $y$-axis. It also shows the evolution of $C$ in time, which is the same for all three simulations. The black diamond on the hysteresis curve for $\tau=0$ (blue, dashed line) marks the simulation-time $t = 30000$. 
     This is marked in the inset by gray dotted lines.}
    \label{fig:hysteresis}
\end{figure*}

Recently, hysteresis in the turbulence to zonal flow transition has been shown in a numerical simulation of the modified Hasegawa-Wakatani model in Ref. \citenum{guillon2025phase}. It was concluded therein that this behavior can be described as a phase transition\cite{guillon2025flux, guillon2025self}. Similar observations were described in a related context before\cite{kim2011intrinsic, gravier2017transport}. In Figure \ref{fig:hysteresis} we present a reproduction of this experiment and additionally investigate \textsc{flr} effects on the hysteresis. The non-linear behavior of zonal flows, especially the fact that mergers are possible, seem to speak against the term phase transition in this context.

All three simulations start with a Gaussian blob and run for $1000 L_\perp/c_0$ to achieve nonlinear equilibrium in a turbulent state. Then the resistive coupling parameter $C$ is slowly ramped up and down again. A hysteresis loop in the zonal energy $E_{zf} \equiv \frac{1}{2} \left( \partial_x \langle \phi_i \rangle_y \right)^2$ is visible for zero and finite values of $\tau$.

We observe that values of $\tau$ shift the transition from the turbulent to the zonal flow state towards larger values of $C$. Also the maximal zonal energy is increased for larger values of $\tau$.
Note that a sudden rise in zonal energy (marked in Figure \ref{fig:hysteresis}) coincides with the merger that is visible in the zonal flow velocity profile in the inset. This considerably enlarges the volume of the hysteresis loop for this specific simulation. No mergers occurred in the other two simulations presented in Figure \ref{fig:hysteresis}.
That observation once more emphasizes the fact, that unpredictable, nonlinear behavior is important in zonal flow dynamics. 

Another experiment (not shown here) involved two simulations with a zonal flow state evolving from turbulence ($C=0.3, \tau = 0$). After a phase of constant $C$ up to $ t = 1000 L_\perp/c_0$, the parallel coupling parameter $C$ was slowly ramped down ($\Delta C/\Delta t = -0.01 /1000$) to $C=0.15$, which is a turbulent state according to the bifurcation diagram\cite{grander2024hysteresis}. In one case, the simulation stayed in a 
zonal flow state with high $E_{zf}$ up to $t = 2 \times 10^5 L_\perp/c_0$. In the other case, the zonal flows were destroyed at approximately $t = 10^4 L_\perp/c_0, C \lesssim 0.2$ and the simulation continued in a turbulent state. Apart from slightly different initial conditions, the simulations were run with identical parameters.

Both, the nonlinear dynamics within the hysteresis loop as well as the chaotic results in sub-critical states can also be used as arguments, that the term phase transition \cite{guillon2025phase} from thermodynamics might not be adequate in this context.

More fundamentally, the Hasegawa–Wakatani system does not seem to admit a free-energy functional whose minima would define competing equilibrium states, as is required for a genuine thermodynamic phase transition\cite{kardar2007}. The Hasegawa-Wakatani system is non-Hamiltonian in the sense, that it is both driven and dissipative. Therefore no free-energy functional can be derived from the Hamiltonian from a variational principle.
However without a well-defined free-energy functional, there can be no equilibrium metastable states separated by an energetic barrier. Therefore, there seems to be no thermodynamic mechanism capable of producing a true first-order transition.

In that strict sense the hysteresis seems to reflect the coexistence of more than one dynamically sustained attractor in a non-equilibrium system, rather than the crossing of free-energy minima. 

\section{Conclusion}

    We have shown that nonlinear terms in the gyrofluid Hasegawa-Wakatani model lead to 1) a chaotic development of initial zonal flow patterns and 2) chaotic dynamics of seemingly stable zonal flow dominated systems through the merging of zonal flows. The merging of zonal flows is explained with nonlinear local momentum transfer through Reynolds stress. The sensitive dependence on initial conditions seems to suggest that time-averages of single simulations might not be enough to predict e.g. transport coefficient in real live systems. Rather the averages of ensembles of simulations with variations in the initial conditions might be preferable. 

    Furthermore we investigated \textsc{flr} effects on zonal flows as well as their dependency on the parallel coupling parameter $C$. This was done by sampling ensembles of dominant zonal flow mode numbers $q_r^d$ for a range of $\tau$ and $C$. Finite values of the ion to electron temperature ratio $\tau$ significantly alter the dominant zonal flow modes with a simple model $q_r^d(\tau) \propto 1/(1+\tau)$, whereas the connection to the parallel coupling parameter $C$ is not as conclusive.
    
    Moreover we could reproduce previous results from Ref. \citenum{guillon2025phase} concerning the hysteresis in the turbulence to zonal flow transition. This observation could be extended to finite values of $\tau$. Arguments against the appropriateness of calling this bifurcation a phase transition in the strict thermodynamic sense were given.

    The authors are well aware, that the 2D, isothermal, electrostatic, delta-f model presented in this publication cannot be used to predict the behavior of magnetized plasma in fusion experiments. zonal flows apparently do not play such a dominant role in 3D electromagnetic full-f simulations of tokamak-edge plasmas\cite{scott2003computation, kendl2005shear}.
    
    Nevertheless the chaotic developement of initial zonal flow patterns as well as a chaotic dynamical behavior in seemingly saturated states might also be given in higher fidelity models. However, the question of the relevance of the phenomena reported in this publication has to be further investigated, using higher fidelity models (especially 3D).

\begin{acknowledgments}
This research (F.G) was supported by the Austrian Science Fund (FWF) project 10.55776/P33369. F.G. also acknowledges financial support by KKKÖ of ÖAW. 

This work has been partly carried out within the framework of the EUROfusion Consortium, funded by the European Union via the Euratom Research and Training Programme
(Grant Agreement No. 101052200 — EUROfusion) (F.G., F.F.L.). Views and opinions expressed are however those of the
author(s) only and do not necessarily reflect those of the European Union or the European Commission.
Neither the European Union nor the European Commission can be held responsible for them.

The authors thank Cai Hallstrom (Bethel University, Minnesota) for contributions to the code and helpful discussions concerning the zonal flow momentum conservation equation. They also thank Wolfgang Jais for support on the department's gpu cluster.
\end{acknowledgments}

\section*{Author declarations}
\subsection*{Conflict of interest}
The authors have no conflicts to disclose.

\subsection*{Autor contributions}
    \noindent
    \textbf{Fabian Grander:} Data curation (lead); Formal analysis (lead); Investigation (lead); Methodology (equal); Software (supporting); Visualization (lead); Writing - original draft (lead); Writing - revision (lead).
    \textbf{Tobias Gröfler:} Software (lead).
    \textbf{Franz-Ferdinand Locker:} Formal analysis (supporting); Investigation (supporting); Methodology (equal); Writing - original draft (supporting); Writing - revision (supporting).
    \textbf{Manuel Rinner:} Formal analysis (supporting); Investigation (supporting); Writing - original draft (supporting).
    \textbf{Alexander Kendl:} Funding acquisition (lead); Resources (lead); Project administration (lead); Software (supporting); Investigation (supporting); Methodology (equal);  Writing - original draft (supporting); Writing - revision (supporting)

\section*{Data availability}
The data that support the findings of this study are available from the corresponding author upon reasonable request.

\bibliography{zfstability}

\end{document}